# Determination of the spin axis in quantum spin Hall insulator monolayer WTe$_2$

Wenjin Zhao[1]†, Elliott Runburg[1]†, Zaiyao Fei[1], Joshua Mutch[1], Paul Malinowski[1], Bosong Sun[1], Xiong Huang[2,3], Dmytro Pesin[4], Yong-Tao Cui[2,3], Xiaodong Xu[1,5], Jiun-Haw Chu[1], David H. Cobden[1]*

[1]Department of Physics, University of Washington, Seattle WA 98195, USA
[2]Department of Physics and Astronomy, University of California, Riverside, Riverside CA 92521, USA
[3]Department of Materials Science and Engineering, University of California, Riverside, Riverside CA 92521, USA
[4]Department of Physics, University of Virginia, Charlottesville, Virginia 22904, USA
[5]Department of Materials Science and Engineering, University of Washington, Seattle WA 98195, USA
†These authors contributed equally
*Corresponding author: cobden@uw.edu

## ABSTRACT

**Evidence for the quantum spin Hall (QSH) effect has been reported in several experimental systems in the form of approximately quantized edge conductance. However, the most fundamental feature of the QSH effect, spin-momentum locking in the edge channels, has never been demonstrated experimentally. Here, we report clear evidence for spin-momentum locking in the edge channels of monolayer WTe$_2$, thought to be a two-dimensional topological insulator (2D TI). We observe that the edge conductance is controlled by the component of an applied magnetic field perpendicular to a particular axis, which we identify as the spin axis. The axis is the same for all edges, situated in the mirror plane perpendicular to the tungsten chains at $40 \pm 2°$ to the layer normal, implying that the spin-orbit coupling is inherited from the bulk band structure. We show that this finding is consistent with theory if the band-edge orbitals are taken to have like parity. We conclude that this parity assignment is correct and that both edge states and bulk bands in monolayer WTe$_2$ share the same simple spin structure. Combined with other known features of the edge states this establishes spin-momentum locking, and therefore that monolayer WTe$_2$ is truly a natural 2D TI.**

## I. INTRODUCTION

The two-dimensional topological insulator (2D TI) is the original and archetypal topological solid with time-reversal symmetry [1-5]. The nontrivial topology of the electronic states in a 2D TI endows it with gapless one-dimensional edge modes with the unusual property of being helical. In a helical mode the electron's spin is locked to its momentum: it points in a particular direction for electrons moving one way along the edge and in the opposite direction for electrons moving the other way. This causes current flowing along opposite edges to have opposite spin polarization, a phenomenon referred to as the quantum spin Hall (QSH) effect. In addition, in a helical mode elastic backscattering in zero magnetic field is prohibited. In view of this, a single-edge quantized conductance of $e^2/h$ in zero field has been considered evidence for the QSH effect, $e$ being the electron charge and $h$ Planck's constant. However, inelastic and magnetic backscattering can still occur [6-10] and the quantization is not expected to be accurate, unlike in the quantum Hall effect. Indeed, the leading 2D TI candidate materials, HgCdTe [11] and InAs/GaSb quantum wells [12] and monolayer WTe$_2$ [13,14], do not exhibit well quantized conductance. The real essence of the QSH effect is spin-momentum locking in the edges, a property that has never been clearly demonstrated.

In principle, the natural way to determine the existence of a spin axis in a helical mode is to apply a magnetic field ***B*** and vary its orientation. Zeeman coupling of the field to the electron spin will gap the mode, thereby decreasing the conductance, except when $\boldsymbol{B}$ is aligned parallel to the spin axis. Unfortunately, in quantum well systems strong orbital coupling to the field masks this effect [11]. Our key finding is that in monolayer (1L) WTe$_2$ the edge conduction does exhibit the magnetic anisotropy expected for Zeeman-coupled helical modes, in both linear and nonlinear response to voltage bias. Moreover, we find that the spin axis, which we call $\boldsymbol{d_{so}}$, is almost independent of edge orientation, gate voltage, and sample, implying that it is a property of the bulk bands near $E_F$ that is inherited by the edge modes. This conclusion is illustrated schematically in Fig. 1(a). The crystal structure of 1L WTe$_2$ is shown in Fig. 1(b), along with its three spatial symmetries: a 2-fold screw axis ($C_{2x}$) along the tungsten chains; a mirror plane ($M_x$) perpendicular to the chains (the *y-z* plane); and inversion (parity *P*). The spin axis $\boldsymbol{d_{so}}$ is in the mirror plane, making an angle $\varphi_{so} = 40 \pm 2°$ with the *z*-axis. We will also show that this is consistent with a $\boldsymbol{k} \cdot \boldsymbol{p}$ analysis of the bands using appropriate symmetries of the orbitals [15-18].

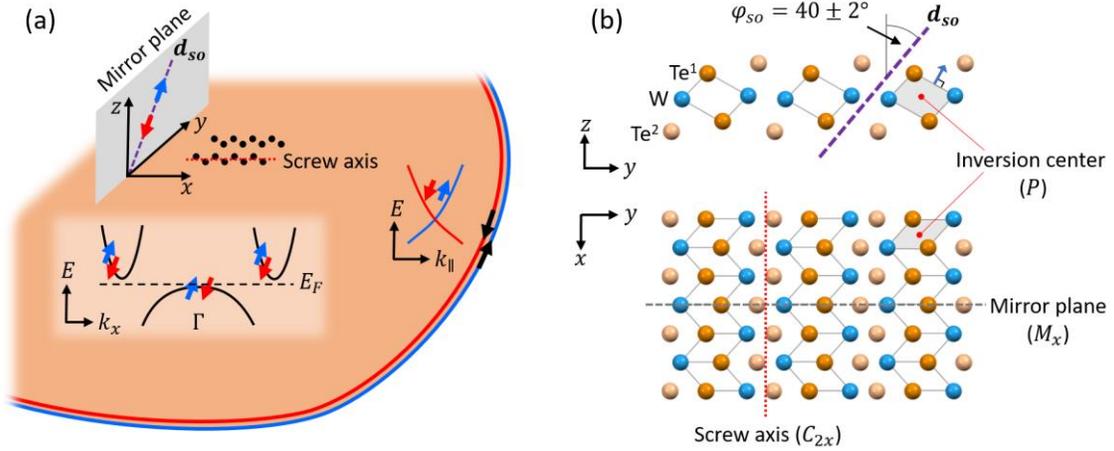

FIG. 1. The spin axis of monolayer WTe$_2$. (a) Cartoon illustrating our key findings: the edge modes on monolayer WTe$_2$ are spin polarized along an axis $\boldsymbol{d_{so}}$ that is independent of edge orientation and shared with the bulk conduction and valence band edges, sketched here schematically near Γ. (b) Side and top views of the 1T′ monolayer structure indicating the three spatial symmetries (*P*, $M_x$ and $C_{2x}$), and showing the direction of $\boldsymbol{d_{so}}$, which is in the *y-z* mirror plane making angle $\varphi_{so} = 40 \pm 2°$ with the *z*-axis. Tungsten atoms are shown blue, and the two inequivalent tellurium atoms, Te$^1$ and Te$^2$, are different shades of orange. The screw axis (red dotted line) is along the center of a zigzag tungsten chain.

## II. RESULTS

Figure 2 illustrates how the conductance depends on the orientation of $\boldsymbol{B}$ for one device, MW5. An optical image of this device is shown in Fig. 2(a). The monolayer WTe$_2$ flake overlies six Pt contacts, encapsulated between hexagonal boron nitride dielectric layers with a graphite gate electrode beneath. Polarized Raman spectroscopy (inset to Fig. 2(a)) was used to determine the orientation of the *x*-axis (see fig. S3) [19]. The temperature and gate voltage ($T = 4$ K, $V_g = -2.7$ V) were chosen such that the bulk is insulating and edge conduction dominates [13,14]. Microwave impedance microscopy (MIM) could be used to map the in-plane conductivity and thereby establish the precise geography of the edges (Fig. 2(b)) [20]. Importantly, MIM reveals any cracks in the monolayer, which create internal conducting edges that are hard to detect by other means yet are critical for electrical transport.

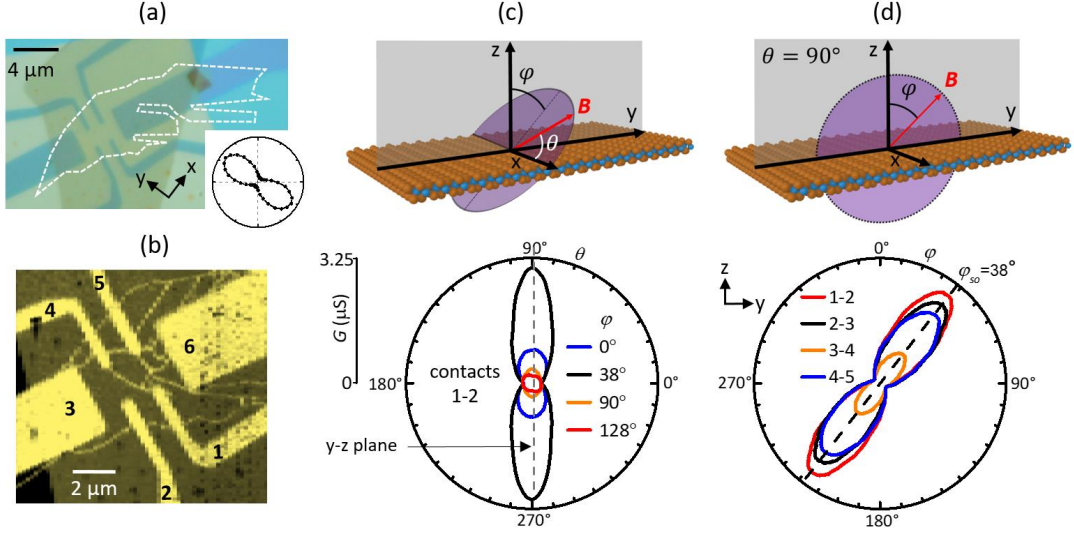

FIG. 2. Sensitivity of edge conduction in monolayer WTe$_2$ to magnetic field orientation. (a) Optical microscope image of device MW5. The monolayer WTe$_2$ flake is outlined with a white dashed line. Inset: polar plot of the polarized Raman $A_1$ mode intensity used to determine the *x* and *y* crystal axes. (b) Image of the device obtained using microwave impedance microscopy to map the local conductivity ($T = 4$ K, $V_g = 0$, $B = 0$), revealing edges and cracks as thin bright lines. (c) Polar plot of conductance $G$ between contacts 1-2 as the magnetic field $\boldsymbol{B}$ (of strength $B = 3$ T) is rotated in the manner indicated in the sketch above ($T = 4$ K, $V_g = -2.7$ V). The sketch also serves to define the angles $\theta$ and $\varphi$ specifying the orientation of $\boldsymbol{B}$ relative to the crystal axes. (d) Polar plot of $G$ for rotation of $\boldsymbol{B}$ in the *y-z* (mirror) plane, as indicated the sketch above. Here measurements are plotted for four contact pairs. For each, $G$ is normalized by its minimum value. In all cases the axis of anisotropy (dashed line) is close to $\varphi = 38°$.

The sketch in Fig. 2(c) shows the coordinates used for the field direction: polar angle $\theta$ relative to the *x*-axis and azimuthal angle $\varphi$ about this axis, with $\varphi = 0$ along the *z*-axis. Below is a polar plot of the conductance $G$ between contacts 1-2 versus $\theta$ at selected values of $\varphi$. We let $\theta$ run from 0° to 360° to parameterize rotating $\boldsymbol{B}$ in a full circle (shaded purple in the sketch). Onsager symmetry requires $G(\boldsymbol{B}) = G(-\boldsymbol{B})$, so the equivalence of the data for $\theta$ and $\theta + 180°$ serves as a check on the experiment. The conductance is maximal in a particular direction, $\boldsymbol{d_{so}}$, which is at $\varphi = 38° \stackrel{\text{def}}{=} \varphi_{so}$ and in the *y-z* plane ($\theta = 90°$). In the plane perpendicular to $\boldsymbol{d_{so}}$, specified by $\varphi = 128°$, the conductance is much smaller and relatively independent of $\theta$ (red trace). There is a high degree of symmetry about the *y-z* (mirror) plane. Figure 2(d) shows the dependence on $\varphi$ within the *y-z* plane for the same contact pair as well as for three others. For every pair the conductance maximum is within a degree of 38° (dashed line), even though the edges connecting them have different shapes and orientations with respect to crystal axes as shown by fig. 2(a)&(b). Across five monolayer devices, even though the conductance at zero magnetic field varies, all contact pairs show similar behavior with $\varphi_{so}$ in the range $40 \pm 2°$ (see figs. S4 to S6), indicating that disorder on the edges only weakly affects the magnetic anisotropy of edge conduction. Devices made from bilayer WTe$_2$, which is topologically trivial, did not show this behavior (fig. S13).

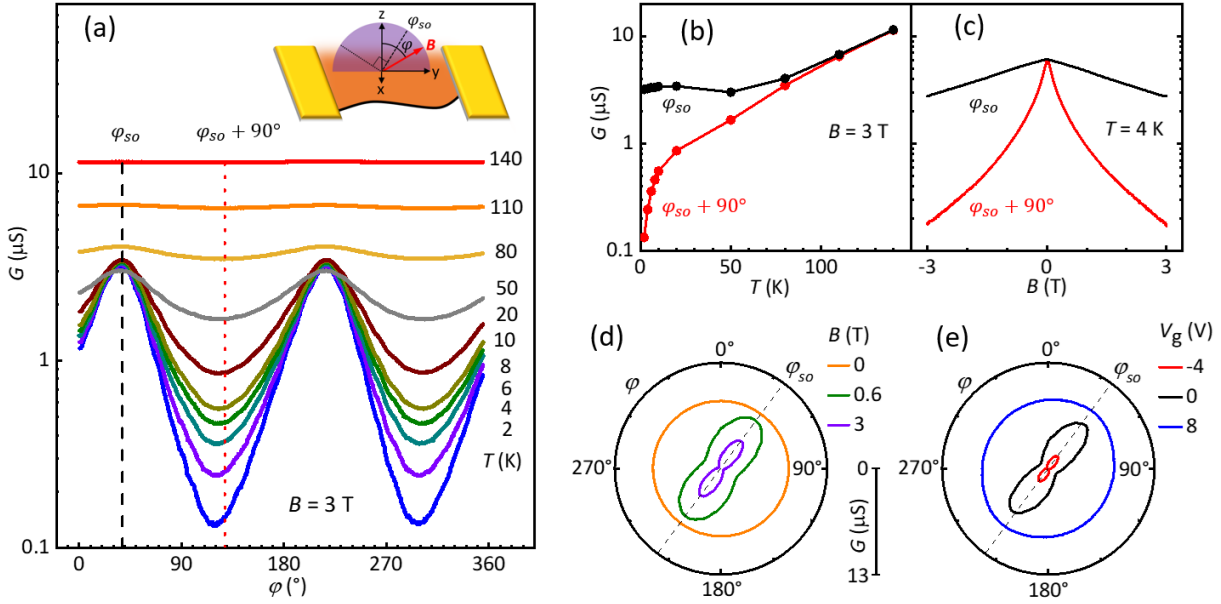

FIG. 3. Dependence of the anisotropy on temperature, field strength and gate voltage. (All measurements are for contacts 1-2 in device MW5). (a) Two-terminal conductance $G$ measured while rotating $\boldsymbol{B}$ in the $y$-$z$ plane (see sketch above), at a series of temperatures ($B = 3$ T; $V_g = -2.7$ V). (b) Corresponding temperature dependence of $G$ at fixed field orientations $\varphi = \varphi_{so}$ and $\varphi = \varphi_{so} + 90°$. (c) Effect of sweeping the field at $T = 4$ K for the same two orientations. (d) Polar plots of $G$ versus $\varphi$ for different field strengths ($V_g = -2.7$ V, $T = 4$ K). (e) Similar plots for different $V_g$ ($B = 3$ T, $T = 4$ K). (At $V_g = +8$ V, values are divided by 3).

Figure 3(a) shows the conductance measured as $\boldsymbol{B}$ is rotated in the $y$-$z$ plane at a series of temperatures. At $T = 140$ K, bulk conduction dominates and the effect of the field is very weak. On cooling below ~100 K the bulk conduction freezes out and $G$ depends increasingly strongly on $\varphi$. Figure 3(b) compares the conductance for $\varphi = \varphi_{so}$, where the temperature dependence is weakest, with that for $\varphi = \varphi_{so} + 90°$. Figure 3(c) shows how $G$ depends on the field strength at 4 K: it drops much faster for $\varphi = \varphi_{so} + 90°$ than for $\varphi = \varphi_{so}$. Figure 3(d) shows that $\varphi_{so}$ is independent of $B$, and Fig. 3(e) shows that it is almost independent of $V_g$, varying by less than 2° (see fig. S7).

While it is still not known exactly which mechanisms dominate the resistivity and magnetoresistance of helical quantum wires (see Refs [6-10] and references therein), it is clear that the existence of a specific orientation in which $\boldsymbol{B}$ has minimal effect implies that the spins at the Fermi level $E_F$ are aligned in this direction. Moreover, the constancy of $\boldsymbol{d_{so}}$ has a natural interpretation based on the symmetry properties of 1L WTe$_2$ (see Fig. 1). The low-energy physics of bulk monolayer WTe$_2$ is captured by an effective $\boldsymbol{k} \cdot \boldsymbol{p}$ model constructed around the $\Gamma$-point in the Brillouin zone. The minimal model contains four bands, built from two orbitals (the bottom of the conduction band and the top of the valence band at $\Gamma$) and two spin states. Each orbital must have definite symmetry (even or odd) with respect $\mathcal{P}$, $M_x$, and $C_{2x}$ (see Fig. 1(b)). One group of first-principles studies [5,21-24] assumed that the two orbitals have opposite $P$ and the same $M_x$ symmetry (implying opposite symmetry under $C_{2x}$). In this case, the matrix elements of the $\boldsymbol{k} \cdot \boldsymbol{p}$ Hamiltonian that connect these basis states must be even under $M_x$ and time-reversal $\mathcal{T}$, and odd under $\mathcal{P}$ (and $C_{2x}$) transformations. The leading-order allowed SOC terms then have the form $H_{SOC}^I = \tau_x(v_x p_y \sigma_x + v_y p_x \sigma_y + v_z p_x \sigma_z)$, where the Pauli matrices $\tau_i$ ($\sigma_i$) act in the orbital (spin) space, and coefficients $v_i$ have dimensions of velocity. SOC of this form projected into edge modes will produce an edge- and momentum-dependent spin axis that is not in the $y$-$z$ mirror plane. This is not consistent with our findings.

Another group of first-principles studies [15-18] assumed instead that the orbitals have the same $\mathcal{P}$ but opposite $M_x$ (and $C_{2x}$) symmetry. In this case the matrix elements connecting them must be even under $\mathcal{T}$ and $\mathcal{P}$, and odd under $M_x$ (and $C_{2x}$). The leading-order SOC terms then have the form

$$H_{SOC}^{II} = \tau_y(\lambda_y\sigma_y + \lambda_z\sigma_z) \equiv \tau_y \boldsymbol{D}\cdot\boldsymbol{\sigma},$$

where coefficients $\lambda_y$ and $\lambda_z$ have dimensions of energy and $\boldsymbol{D} \equiv (0, \lambda_y, \lambda_z)$ can be thought of as a vector in the y-z plane making an angle of $\varphi_D = \tan^{-1}|\lambda_y/\lambda_z|$ with the z-axis. The bulk bands near Γ then consist of pairs related by the time reversal, whose spin expectation values are either along or opposite to $\boldsymbol{D}$ after tracing out the orbital components. Edge states near $E_F$ will carry this same spin polarization as long as the coupling to bands far from $E_F$ can be ignored. The presence of confining electric fields at the edge does not change this conclusion: even though they can modify the edge dispersion, their contribution to the spin-orbit coupling is small compared with that of the electric fields in the atomic orbitals that govern the bulk SOC. This is consistent with our findings if we identify $\boldsymbol{D}$ with $\boldsymbol{d}_{so}$, and hence $\varphi_D$ with $\varphi_{so} = 40 \pm 2°$. Strictly speaking, the maximum conductance is expected when the direction of $\overleftrightarrow{g}\boldsymbol{B}$, not $\boldsymbol{B}$, coincides with the spin polarization direction. For a generic edge, the g-tensor $\overleftrightarrow{g}$ is not constrained by any symmetry and furthermore can depend on details of the edge. This can explain why the angular separation of the minima and maxima in $G$ is not exactly 90°, but ranges down to 83° (in Fig. 3(a) it is 85°.) However, the lack of variation of $\boldsymbol{d}_{so}$ implies either that $\overleftrightarrow{g}$ is nearly isotropic or, at least, that $\boldsymbol{d}_{so}$ remains close to a principal axis of $\overleftrightarrow{g}$ irrespective of the edge structure.

The form of $H_{SOC}^{II}$ can also be derived in a microscopic 8-band tight-binding model [15,18], in which it corresponds to the strongest spin-flip hopping process in the y direction along the Te[1]-W bonds (drawn as lines in Fig. 1(b)). It is not possible to calculate $\boldsymbol{D}$ accurately since the parameters in the models are not known well enough. Nevertheless, it is worth noting that the two interfering hopping paths that involve intermediate Te sites on the adjacent Te[1]-W bonds give the standard Haldane-Kane-Mele [1,25] contribution to $\boldsymbol{D}$. This turns out to be perpendicular to the Te[1]-W bond, in the mirror plane at 29° from the z-axis (indicated with a blue arrow in Fig. 1(b)), which is not too different from the measured value of $\varphi_{so}$ (fig. S8). Near the bulk band edges, which occur at finite values of $p_x$, and zero $p_y$, the two types of the spin-orbit coupling described above are similar to each other, and lead to a quantum spin Hall state with a canted spin quantization axis. The properties of this state are described in Ref. [24].

Helical edge modes are also expected to exhibit a characteristic nonlinear magnetotransport effect [26-28]: In the Taylor expansion of the current-voltage relation, $I(V, \boldsymbol{B}) = G(\boldsymbol{B})V + \gamma(\boldsymbol{B})V^2 + \cdots$, the coefficient $\gamma$ describes nonreciprocal conduction, allowed because the edge breaks inversion symmetry. Its $\boldsymbol{B}$-odd part, $\gamma_a(\boldsymbol{B}) = [\gamma(\boldsymbol{B}) - \gamma(-\boldsymbol{B})]/2$, is sensitive to the edge spin texture. Two main mechanisms contribute to $\gamma_a(\boldsymbol{B})$ at low temperature [29]. In one, an exchange field proportional to the current-induced spin polarization (CISP) [30], which is along $\boldsymbol{d}_{so}$, adds to the applied field to modify the conductance. In the other, nonlinear dispersion and broken inversion symmetry in the edge energy spectrum together lead to a lack of cancellation of quadratic-in-voltage currents carried by the two helical branches. For both mechanisms, $\gamma_a(\boldsymbol{B})$ should vanish under the same condition that the suppression of $G$ is maximal.

Motivated by this prediction, we measure $\gamma_a$ by applying an a.c. bias, $V_f$, at frequency $f$ and measuring the second-harmonic component, $I_{2f}$, of the current at a fixed phase with $V_f$ chosen such that $|I_{2f}| \ll I_f$, the first-harmonic component. The dependence of $\gamma = 2I_{2f}/V_f^2$ on field amplitude along a particular axis at a given $V_g$ typically has a heart-beat shape, as illustrated in the upper inset to Fig. 4 (see figs. S9 and S10). The upper part of Fig. 4 shows an example of how $\gamma_a$ varies as $\boldsymbol{B}$ is rotated in the y-z plane at several temperatures. The lower part of the figure shows $G = I_f/V_f$ measured simultaneously. As $T$ is decreased, $\gamma_a$ grows in concert with the anisotropy of $G$. (We note that in the bilayer device, where the anisotropy remained very small, $\gamma_a$ was unmeasurably small). Just as predicted, $\gamma_a$ passes through zero at a temperature-independent value of $\varphi$ very close to the minimum in $G$, at ~120° (blue dotted line) (see fig. S11). In fact, the CISP mechanism alone quantitatively reproduces the full angular dependence of $\gamma_a$ when $B$ is not too small, and the single fitting parameter provides an estimate of the strength of electron-electron interactions in the helical edge (see fig. S12) [29].

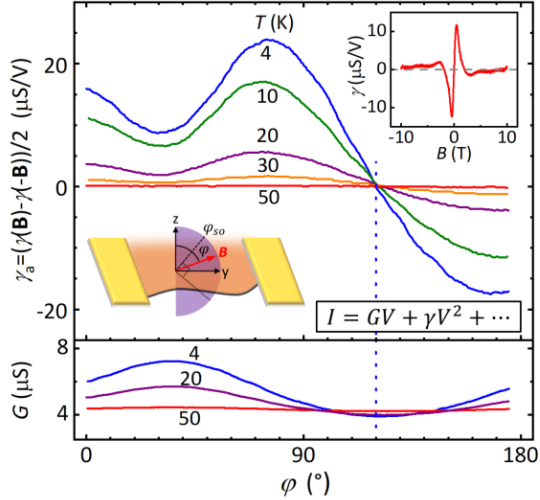

**Fig. 4. Anisotropy of the nonlinear conductance.** Simultaneous measurements of the field-antisymmetrized nonlinear coefficient $\gamma_a$ (top) and linear conductance $G$ (bottom) made as ***B*** is rotated in the *y-z* plane at selected temperatures (contacts 1-2 in device MW5, $V_g = -2.7$ V, a.c. bias 15 mV at $f = 101$ Hz). The field strength $B = 0.3$ T was chosen to maximize $\gamma$. Inset: example of the variation of $\gamma$ with $B$ (see fig. S10).

## III. CONCLUSION

Since the original theoretical prediction that 1L WTe$_2$ could be a topological insulator [5], it has already been established experimentally [13,14,31] that the edge states span the bulk band gap and have a conductance approaching $e^2/h$ on a length scale of 100 nm which indicates suppressed backscattering from what is likely heavy disorder on the sample edge. Our demonstration of the spin polarization and the behavior of $\gamma_a$ characteristic of a helical mode removes any doubt that these edge states are helical. This cements the identification of 1L WTe$_2$ as a natural two-dimensional topological insulator. In addition, the surprisingly simple spin texture, described by a single spin axis common to the edge modes and bulk bands, has implications for superconductivity [32,33]; and, taken together with the large bulk transport gap (~50 meV) and the possibilities afforded by layering together with other materials including magnets [28] and superconductors [34], creates a uniquely powerful platform for topological and spin-related physics and applications [35].

## IV. METHODS

### A. Device fabrication

hBN crystals were mechanically exfoliated onto thermally grown SiO2 on a highly doped Si substrate. The thickness of hBN flakes used as top and bottom dielectrics are listed in Table S1. For the devices which have a bottom gate, the few-layer graphite is covered by an hBN flake (bottom hBN) by using a polymer-based dry transfer technique. After dissolving the polymer, the hBN/graphite was annealed at 400 °C for 2 h. Next, Pt metal contacts were deposited at ~7 nm thickness on the lower hBN or hBN/graphite by standard e-beam lithography and metallized in an e-beam evaporator followed by acetone lift-off. Then the Pt contacts on hBN or hBN/graphite were annealed at 200 °C for 1 h. WTe2 crystals were exfoliated in a glove box (O2 and H2O concentrations < 0.5 ppm). The monolayer or bilayer WTe2 flake was picked up under another hBN flake (top hBN) or graphite/hBN stack. The stack was then put down on the Pt contacts in the glove box. Finally, another step of e-beam lithography and metallization (Au/V) was used to define wire-bounding pads connecting to the metal contacts and the graphite gate.

### B. Electrical measurements

Electrical measurements were carried out in a PPMS DynaCool cryostat (Quantum Design, Inc.) with a base temperature of 2 K and magnetic field up to 14 T and an Oxford He-4 VTI cryostat with temperature down to 1.6 K and magnetic field up to 14 T. A 1mV a.c. excitation at 101 Hz was applied for linear responses. For second-harmonic responses, a 15 mV a.c. excitation at 101 Hz was applied while a 30 µF capacitor was connected in series with the device. The device in the PPMS cryostat can be rotated along two axes. One axis rotator is supplied by the PPMS cryostat and another axis rotator was bought from attocube and is mounted on the chip carrier.

### C. Polarized Raman measurements

Raman spectroscopy was performed in vacuum at room temperature. He-Ne laser light at 632.8 nm was focused down to a spot size of ~2 µm by an objective at normal incidence. BragNotchTM filters were used to clean the incident and suppress the Rayleigh scattering. The reflected light was analyzed by an Andor spectrometer with a 1200 groove/mm diffraction grating. For thick $WTe_2$, a laser power of 1 mW and an integration time of 30 seconds was adopted, while monolayer flakes were measured at 150 µW and integrated for 3 minutes. In the polarization dependence measurements (colinear geometry), a linear polarizer and a half-wave plate (HWP) were placed right before the objective, shared by the incident and reflected lights. To determine the crystal axes of monolayer flakes, we used an on-chip needle-shaped thick flake as a reference, where the tungsten chain is known to be along the needle. Based on the relative angle between the polarization patterns of the thick and monolayer flake, we can derive the crystal axes of the monolayer.

### ACKNOWLEDGEMENTS

The authors acknowledge Anton Akhmerov, Leonid Glazman, Cenke Xu, Lukasz Fidkowski, Alexander Lau, Lukas Muehler, Likun Shi, Justin Song, Boris Spivak, Ronny Thomale, Mark Rudner, and Di Xiao for insightful discussions. The experiments and analysis were supported by NSF DMR grants MRSEC 1719797 and EAGER 1936697. The work of DP was supported by the National Science Foundation grant No. DMR-1853048. Materials synthesis at UW was partially supported by the Gordon and Betty Moore Foundation's EPiQS Initiative, Grant GBMF6759 to JHC. XH and YTC. acknowledge support from the Hellman Fellowship award and the seed fund from SHINES, an EFRC funded by DOE BES under award number SC0012670.

Supplementary Information for

# Determination of the spin axis in quantum spin Hall insulator monolayer WTe$_2$

Wenjin Zhao†, Elliott Runburg†, Zaiyao Fei, Joshua Mutch, Paul Malinowski, Bosong Sun, Xiong Huang, Dmytro Pesin, Yong-Tao Cui, Xiaodong Xu, Jiun-Haw Chu, David H. Cobden*

†These authors contributed equally
*Corresponding author: cobden@uw.edu

SI I. Preparation and characterization of WTe$_2$ devices

We measured five monolayer WTe$_2$ devices (MW3, MW5, MW6, MW7, MW8) and one bilayer WTe$_2$ device (BW7). Figure S1 shows optical images of some monolayer WTe$_2$ and device MW5.

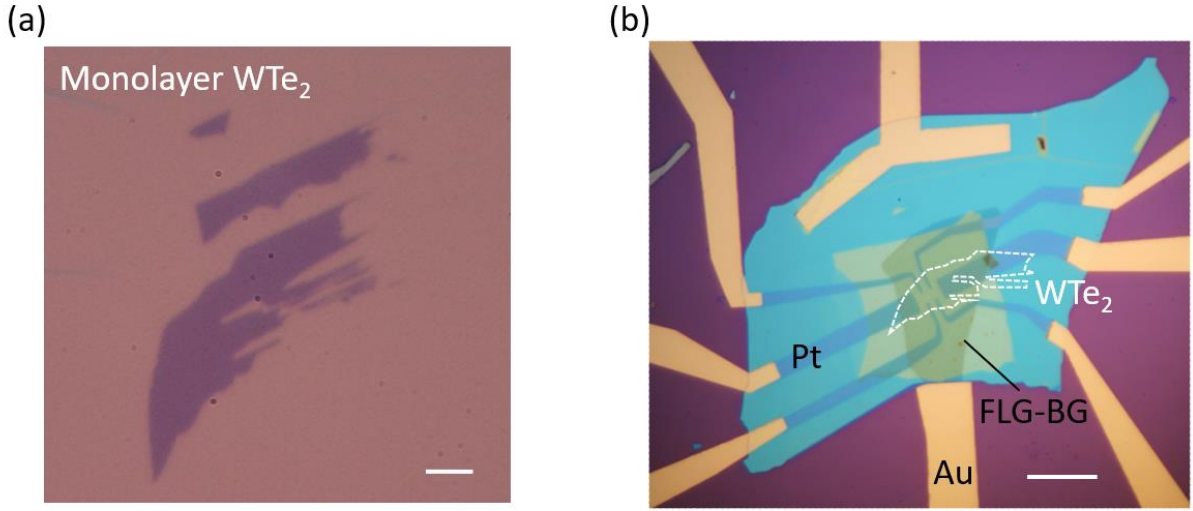

**FIG. S1. Optical image of monolayer WTe$_2$ and device MW5.** (a) Optical microscope image of monolayer WTe$_2$ flakes. Scale bar: 4 µm. (b) Optical microscope image of device MW5. The monolayer WTe$_2$ flake is outlined by a white dashed line. Scale bar: 10 µm.

Table S1 lists the thicknesses of the top and bottom hBN and corresponding areal geometric capacitance $C_g$. The change in electron-hole density imbalance is approximated $n_e = C_g V_g / e$, where $C_g = \epsilon_{hBN}\epsilon_0/d_{hBN}$, where we take $\epsilon_{hBN} = 4$ for the dielectric constant of hBN [36], and $d_{hBN}$ is the thicknesses of the hBN flakes measured by AFM.

| Device label | WTe$_2$ | Top hBN (nm) | Bottom hBN (nm) | $C_g$ (mF/m$^2$) |
|---|---|---|---|---|
| MW3 | Monolayer | 11.4 | 14 | 2.53 |
| MW5 | Monolayer | 16 | 36 | 0.98 |
| MW6 | Monolayer | 8 | 22 | 1.61 |
| MW7 | Monolayer | 12 | 8 | 4.43 |
| MW8 | Monolayer | 22 | 17 | 2.08 |
| BW7 | Bilayer | 8 | 25 | 1.42 |

**Table S1.** Thickness of hBN measured by AFM, and capacitances for bottom gate for devices.

Figure S2 shows gate dependence of two-terminal conductance for device MW3 and MW5 at different temperatures. Because a bump exists in the gate dependence of device MW5, most of the measurements were taken at $V_g = -2.7$ V to avoid it.

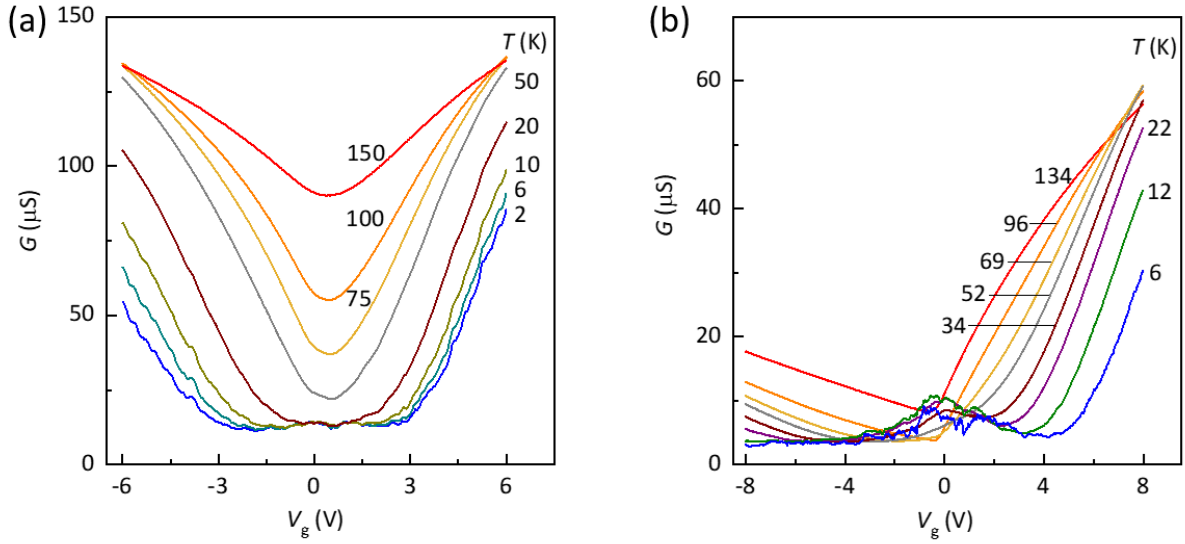

**FIG. S2.** Gate dependence of the linear conductance between two adjacent contacts at different temperature for device MW3 (panel (a)) and contacts 1-2 in device MW5 (panel (b)).

SI II. Determination of crystal axes of monolayer WTe$_2$

After the fabrication process, the crystal axes were determined by polarized Raman measurements at room temperature. The Raman spectra for bulk and monolayer WTe$_2$ are shown is Fig. S3(b) and (e). The peaks located at around $160~\text{cm}^{-1}$ and $210~\text{cm}^{-1}$ are defined as P10 and P11, respectively. For both bulk and monolayer WTe$_2$, the magnitude of Raman intensity for P11 is smallest along the *x*-axis and largest along the *y*-axis.

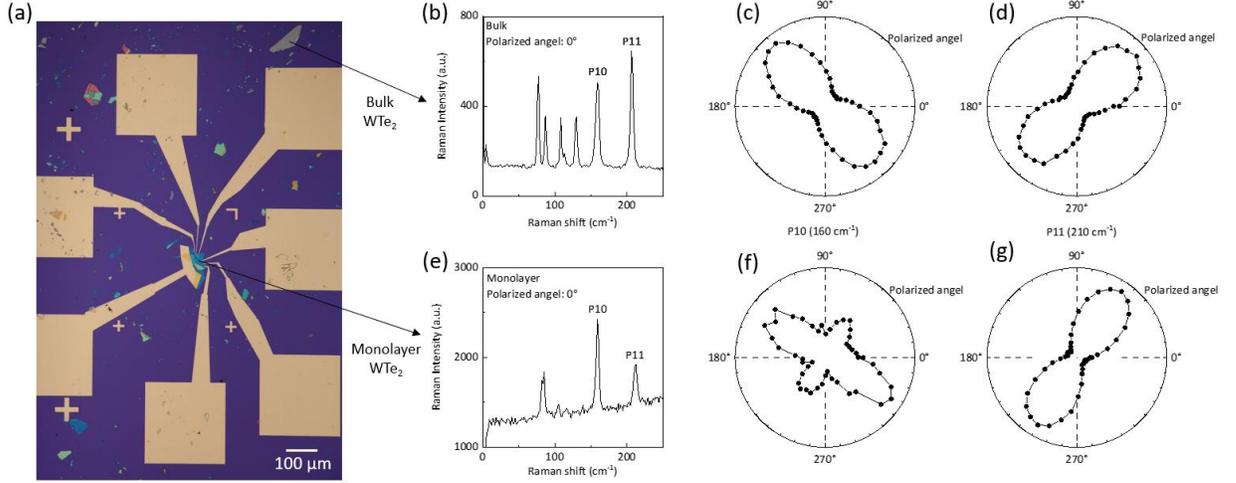

**FIG. S3. Polarized Raman measurements.** (a) Optical picture of device MW7. Scale bar: 100 μm. (b) Raman spectrum for bulk WTe$_2$. (c) and (d) Raman intensity of bulk WTe$_2$ as a function of polarized angle for P10 and P11, respectively. (e) Raman spectrum for monolayer WTe$_2$. (f) and (g) Raman intensity of monolayer WTe$_2$ as a function of polarization angle for P10 and P11, respectively.

SI III. Process of interpolation

In the main text, we define $\theta$ and $\varphi$ as shown in Fig. 2(c). In the experiment, the direction of magnetic field ($\boldsymbol{B}$) was measured in $\alpha$ and $\delta$ coordinates as shown in the Fig. S4(a), where the polar angle $\alpha$ is relative to the *z*-axis, and the azimuthal angle $\delta$ is about this axis, with $\delta = 0$ along the *x*-axis. Figure S4(b) shows the raw data for device MW5 when $\delta$ is stepped and $\alpha$ is swept. We interpolate the raw data and then do 20-point smoothing for each $\varphi$ before polar plotting. The raw data are plotted as magenta lines in Fig. S4(c).

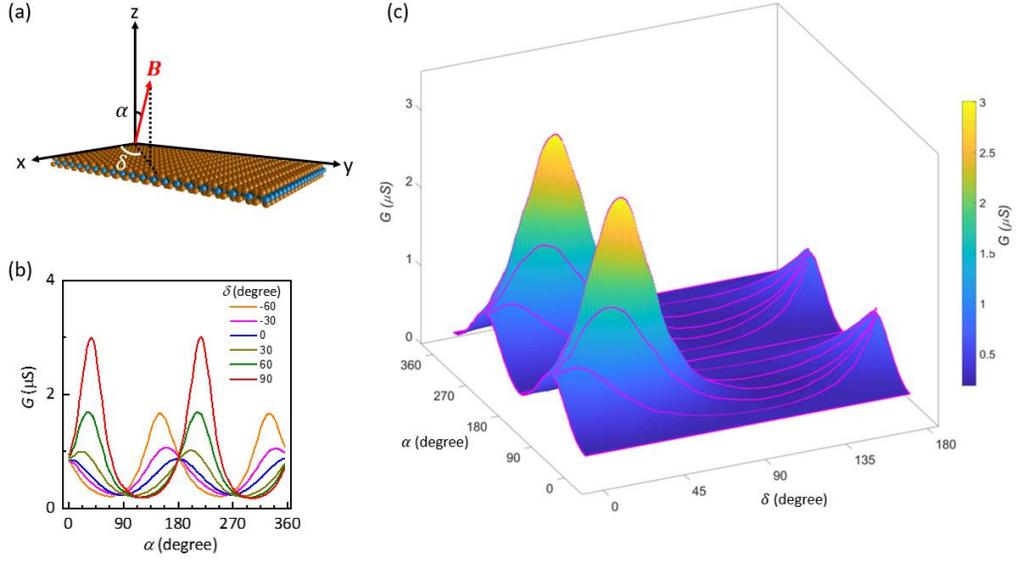

**FIG. S4. Interpolation process.** (a) Definitions of $\alpha$ and $\delta$ relative to the crystal axes of monolayer WTe$_2$ (W, blue and Te, orange). (b) Linear conductance between two adjacent contacts as a function of $\alpha$ at different values of $\delta$ at $B = 3$ T, $T = 4$ K. (c) Color surface plot of interpolated data. The magenta lines are the raw data as shown in panel (b).

For all the monolayer devices, we deduced $\varphi_{so}$ from the raw data first and then used the same interpolation procedure as shown in Fig. S5.

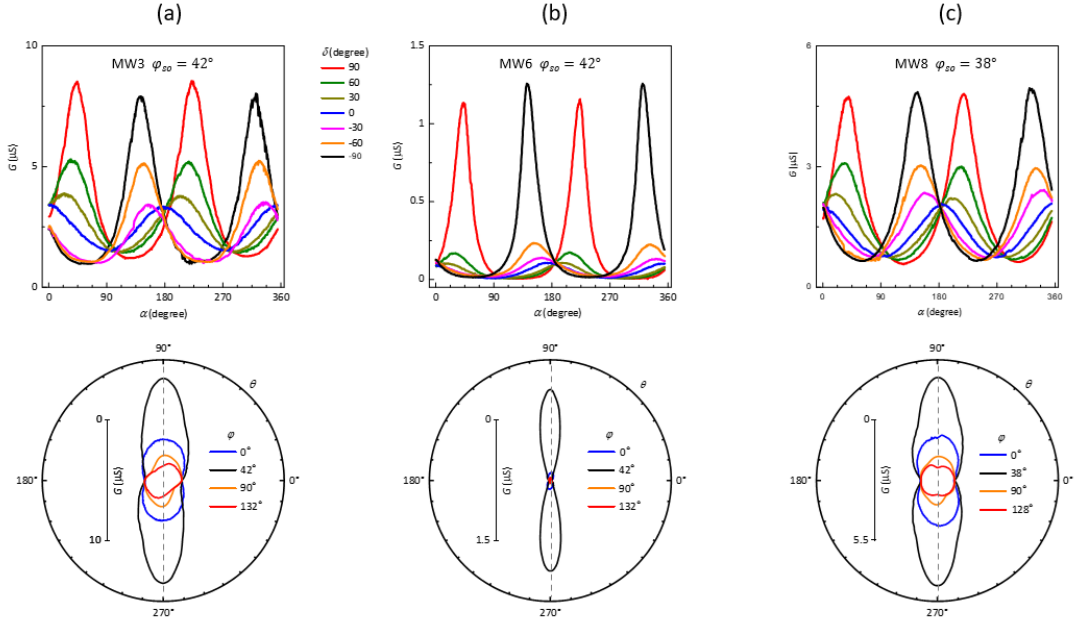

**FIG. S5. Raw data and interpolated data for other monolayer WTe$_2$ devices.** (a) - (c) Linear conductance between two adjacent contacts as presented in $\alpha$ and $\delta$ coordinates (top row) and $\theta$ and $\varphi$ coordinates (bottom row) at $B = 9$ T, $T = 2$ K.

## SI IV. Comparison of $\varphi_{so}$ of different edges

The linear two-terminal conductance ($G$) is maximum at $\varphi = \varphi_{so}, \theta = 90°$ under external magnetic field. As shown is Fig. S5, $\varphi_{so}$ varied among devices with a standard deviation of approximately 2 degrees. This variation could be associated with fabrication differences, contamination, different batches of crystals, and the mounting on the rotator. For a given device, all these factors should be minimal. Figure S6 shows the edge orientations of each monolayer device and $G/G_{\min}$ as a function of $\varphi$ at $\theta = 90°$ (the *y-z* mirror plane). Microwave impedance images are shown when available, otherwise the AFM image is shown instead. As expected, for a given device the standard deviation of $\varphi_{so}$ is less than 1° even for device MW8, which has very many cracks.

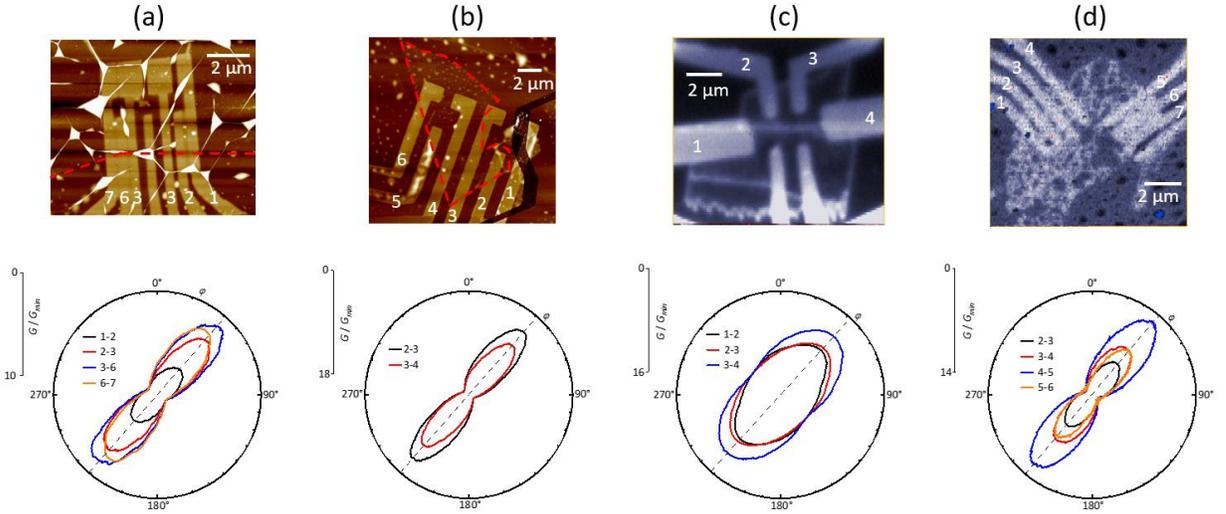

**FIG. S6. Edge orientation and $G/G_{\min}$ in the *y-z* mirror plane.** (a) - (d) Edge orientation (top) and $G/G_{\min}$ (bottom) vs $\varphi$ at $\theta = 90°$ for devices MW3,6,7,8, respectively. In (a) and (b), the monolayer WTe$_2$ flake is outlined by a red dashed line. The contacts are labeled numerically. The magnetic field is 9 T for MW3,6,8 and 3 T for MW7. The temperature is 2 K for devices MW3,6,8, and 4 K for MW7.

## SI V. Dependence of $\varphi_{so}$ on gate voltage

$\varphi_{so}$ depends weakly on gate voltage as shown in Fig. S7. A perpendicular electric field can induce a Rashba effect in the monolayer WTe$_2$. Note that the gate is on top in device MW3 and beneath in MW5. A voltage on a single gate changes both the doping and the perpendicular displacement field. The trends of change of $\varphi_{so}$ are opposite for device MW3 and MW5, suggesting that $\varphi_{so}$ is affected by the displacement field which is opposite in the two cases.

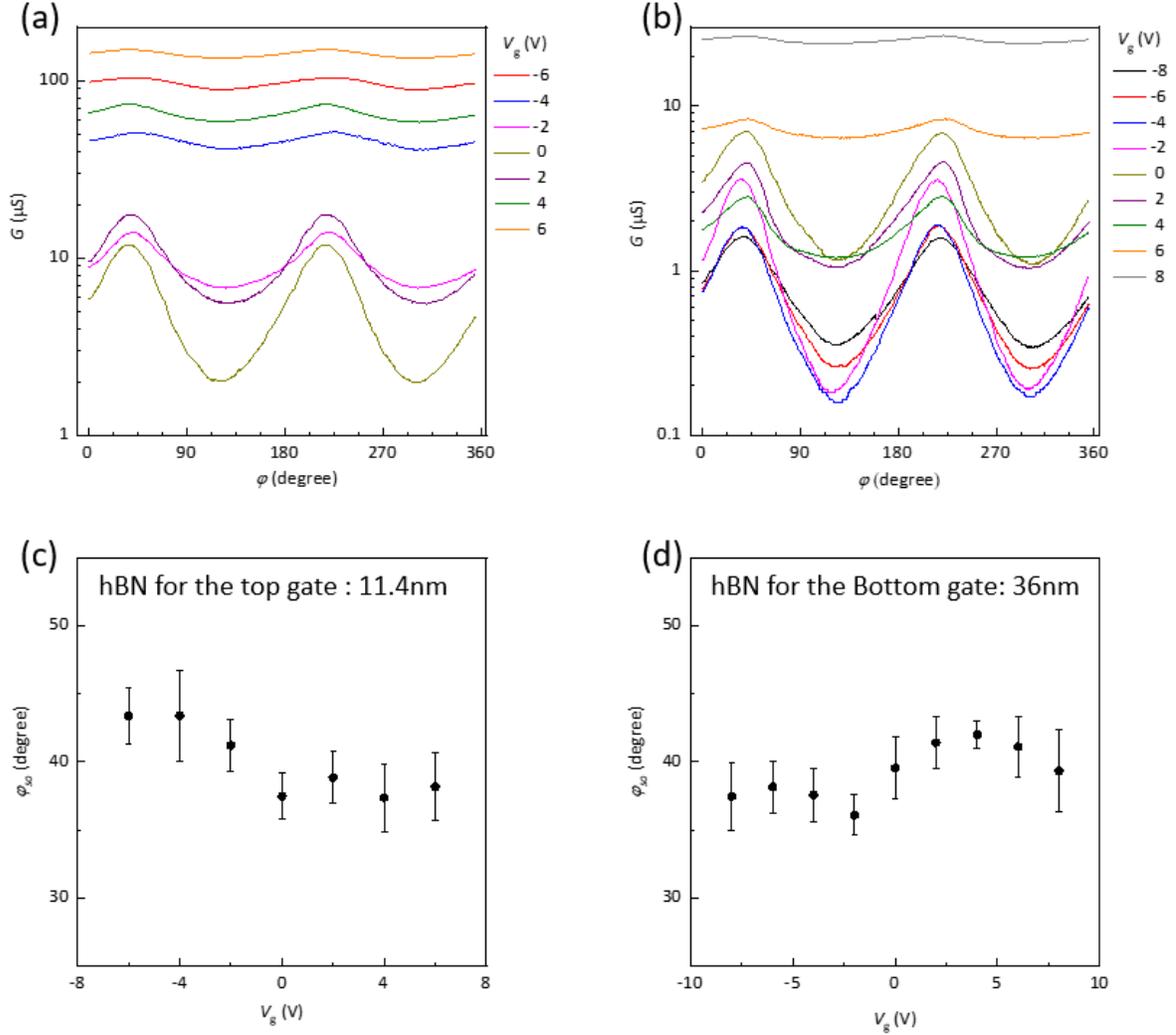

**FIG. S7. Dependence of $\varphi_{so}$ on gate voltage.** (a) and (b) Linear conductance between two adjacent contacts as a function of $\varphi$ in the *y-z* mirror plane for device MW3 and MW5, respectively. $B = 9$ T and $T = 2$ K is used for device MW3. $B = 3$ T and $T = 4$ K is used for device MW5. (c) and (d) $\varphi_{so}$ as a function of gate voltage for devices MW3 and MW5, respectively. The thickness of the gate hBN is labeled in each panel.

SI VI. Estimate of the spin-orbit field direction

"Vertical" hopping between W and Te[1] is forbidden by orbital symmetry (the overlap is zero). However, spin-dependent hopping is allowed, since the "red" paths interfere constructively, as shown in Fig. S8. The amplitude of spin-dependent hopping for each of the red paths must involve $\vec{\sigma} \cdot \vec{d}_1 \times \vec{d}_2$, since the cross product is a pseudovector that plays the role of the "orbital moment" that couples to the spin. The two amplitudes must be subtracted because of the orbital arrangement, since $d_1 - d_2$ hopping mostly happens from the "+" lobe of the d-orbital into the "-" lobe of the p-orbital, and $d'_1 - d'_2$ hopping is between "+" lobes. Hence, the hopping amplitude is proportional to $\vec{\sigma} \cdot \vec{d}_1 \times \vec{d}_2 - \vec{\sigma} \cdot \vec{d}'_1 \times \vec{d}'_2 = \vec{\sigma} \cdot \vec{D}$. As a result, $\vec{D}$ is in the mirror plane and $\varphi_D \approx 29°$.

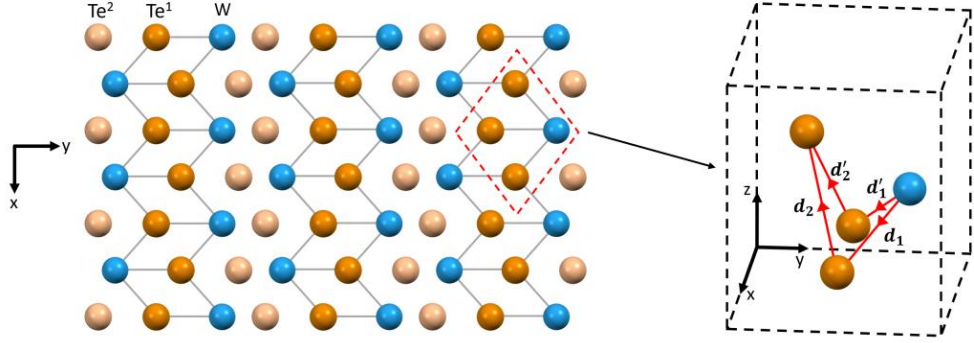

**FIG.S8. Estimate for $\varphi_D$.** Top view of the $1T'$ monolayer WTe$_2$ is shown on the left. A zoom-in view of red dashed area is drawn on the right, where red arrows indicate the two interfering hopping paths.

SI VII. Mechanisms of nonlinear magnetoconductance on a helical edge

As mentioned in the main text, there is a peculiar nonlinear magneto-transport response in monolayer WTe$_2$, which is characterized by a contribution to the current that is odd in magnetic field $\boldsymbol{B}$ and even in the applied voltage $V$. In the Taylor expansion of the $I-V$ characteristic at small voltages, $I(V,\boldsymbol{B}) = G(\boldsymbol{B})V + \gamma(\boldsymbol{B})V^2 + \cdots$, this effect is present when $\gamma(\boldsymbol{B}) = -\gamma(-\boldsymbol{B}) \neq 0$. The origin of this effect can be traced to the Hamiltonian of the helical edge electrons in the presence of a magnetic field. The single-particle disorder-free edge Hamiltonian reads

$$H_{edge} = \epsilon_s(p) + \epsilon_a(p)\boldsymbol{d_{so}} \cdot \boldsymbol{\sigma} + \frac{1}{2}g\mu_B \boldsymbol{B} \cdot \boldsymbol{\sigma} \tag{1}$$

where $p$ is the momentum along the edge, $\epsilon_{s,a}(p) = \pm\epsilon_{s,a}(-p)$, $\mu_B$ is the Bohr magneton, and $g$ is the $g$-factor of the edge electrons. For the time being, we disregard its anisotropy for simplicity. The $g$-tensor will be discussed in a bit more detail below, when we talk about the current-induced polarization on the edge. If we denote the component of $\boldsymbol{B}$ in the direction of $\boldsymbol{d_{so}}$ by $B_\parallel$, and the one perpendicular to $\boldsymbol{d_{so}}$ by $\boldsymbol{B_\perp}$, then the edge spectrum is given by

$$E_\pm(p) = \epsilon_s(p) \pm \sqrt{\left(\epsilon_a(p) + \frac{1}{2}g\mu_B B_\parallel\right)^2 + \left(\frac{1}{2}g\mu_B B_\perp\right)^2} \tag{2}$$

where "$\pm$" pertain to the edge conduction and valence bands.

The sought-for nonlinear response arises from two features of $H_{edge}$. First, it is apparent that $B_\parallel \neq 0$ breaks the "inversion symmetry" in the edge spectrum, $E_\pm(p) \neq E_\pm(-p)$. It is known that such a property in general leads to a current response quadratic in the applied electric field [37], and hence to a nonzero $\gamma(\boldsymbol{B})$. For a helical edge, this response vanishes for a purely linear spectrum $\epsilon_a(p)$; therefore it is sensitive to the band curvature. In general, it is hard to microscopically evaluate the value of $\gamma(\boldsymbol{B})$ that stems from the lack of inversion in the edge spectrum without a reliable model of linear magneto-transport. We will not pursue mechanisms related to the spectrum asymmetry further, other than to point out that the contribution of such mechanisms to the nonlinear magneto-transport vanishes when $B_\parallel = 0$, i.e., $\boldsymbol{B} \perp \boldsymbol{d_{so}}$, which is consistent with our experimental findings. Instead, we will show below that a semi-phenomenological model based on the current-induced spin polarization provides an adequate understanding of the measured nonlinear signal with minimal tuning of parameters.

The second feature of $H_{edge}$ relevant for the nonlinear transport is the edge spin-momentum locking due to a strong spin-orbit coupling and the lack of spatial inversion symmetry on the edge. Under such circumstances, there exists current-induced spin polarization (CISP) of the itinerant carriers. For $H_{edge}$, the spin polarization, $\boldsymbol{s}$, is directed along $\boldsymbol{d_{so}}$. We define it as the difference between the local densities of electrons with spins along or opposite to $\boldsymbol{d_{so}}$ thus: $\boldsymbol{s} = \frac{1}{ev_{edge}} I \boldsymbol{d_{so}}$ (taking $\boldsymbol{d_{so}}$ to be a unit vector). Here $e$ is the electron's charge, and $v_{edge}$ is a coefficient with the dimensions of speed. For a clean non-interacting edge, $v_{edge}$ can be extracted from $H_{edge}$, but in general should be kept as a phenomenological parameter. In the presence of electron-electron interaction, finite spin polarization leads to a self-energy that modifies $H_{edge}$. In the mean-field approximation, and for an $SU(2)$ invariant point-like interaction, the current-induced correction to $H_{edge}$ has the form of an effective Zeeman coupling, and reads [30]

$$\delta H_Z = \frac{1}{2} g_\parallel \mu_B \beta I \boldsymbol{d_{so}} \cdot \boldsymbol{\sigma} \tag{3}$$

In $\delta H_Z$, we introduced $g_\parallel$, the effective g-factor in the direction of $\boldsymbol{d_{so}}$, which defines the Zeeman field for the actual externally applied $B$ field, and a phenomenological coefficient $\beta$ that defines the current-induced effective magnetic field felt by the edge electrons, $\boldsymbol{B_{ex}} = \beta I \boldsymbol{d_{so}}$. The coefficient $\beta$ is related to the electron-electron interaction strength and is discussed below.

At this point we can discuss the influence of the CISP on the transport properties of the edge electrons. In what follows, we assume that one of the principal axes of the effective $g$-tensor of the edge electrons is directed along $\boldsymbol{d_{so}}$. This need not be the case in general, even if the bulk band structure consists of bands fully polarized along and opposite to $\boldsymbol{d_{so}}$, because of the orbital effects associated with the nontrivial geometry of such bands [38]. However, the fact that the directions of the $B$ field that correspond to the minimum and maximum magneto-conductance on the edge are at almost 90° to each other suggests that such orbital effects are weak, and $\boldsymbol{d_{so}}$ is indeed a principal axis of the g-tensor. This is consistent with strong confinement of the edge states in WTe$_2$ [15,31,34]. It then follows that the total Zeeman field felt by the edge electrons corresponds to a certain total magnetic field, which is a simple vector sum of the external field, $\boldsymbol{B}$, and the current-induced effective field, $\boldsymbol{B_{ex}}$. Consequently, the current flow on the edge is described by the following (self-consistent, in principle) equation: $I(t) = G(\boldsymbol{B} + \beta I(t)\boldsymbol{d_{so}})V(t)$, where $G(\boldsymbol{B} + \beta I\boldsymbol{d_{so}})$ is the linear conductance evaluated at the total magnetic field. Under the experimental conditions, the nonlinear current was small compared with the linear one, hence we can assume that the CISP is determined only by the linear-response current. Further, assuming sinusoidal variation of the voltage with time, expanding to linear order in $\beta$, and switching to polar coordinates in the $y$-$z$ mirror plane, we can obtain a relation between the amplitudes of the first- and second-harmonic signals and hence the coefficient $\gamma(\boldsymbol{B})$ in the Taylor expansion of the current as a function of the applied voltage:

$$\gamma(\boldsymbol{B}) = \beta G(\boldsymbol{B}) \left( \frac{\partial G}{\partial B} \cos(\varphi - \varphi_{so}) - \frac{1}{B} \frac{\partial G}{\partial \varphi} \sin(\varphi - \varphi_{so}) \right) \tag{4}$$

This expression for $\gamma(B)$ vanishes for $\boldsymbol{B} \perp \boldsymbol{d}_{so}$, when $\varphi = \varphi_{so} \pm \pi/2$, and $\frac{\partial G}{\partial \varphi} = 0$, since the linear conductance is minimal when $\boldsymbol{B} \perp \boldsymbol{d}_{so}$ at given $B$. Again, this property is observed in the experiment.

We will show that the CISP mechanism explains qualitative features of the angular dependence of $I_{2f}$ (at not too small magnetic fields) (Fig. S12). This fact allows us to estimate the strength of the electron-electron interaction for edge electrons. Assuming long-range Coulomb repulsion, we note that the effective one-dimensional interaction has weak logarithmic dependence on the momentum transfer, $V_{1D}(q) \approx \frac{e^2}{4\pi\epsilon_0\epsilon} \ln\left(\frac{1}{q^2 a^2}\right)$, where $a$ is a cut-off length scale of the order of the confinement length of the edge states, and $\epsilon \approx 4$ is the dielectric constant of the hBN surrounding the WTe$_2$. In real space, one can treat such an interaction as an approximately local one with strength $V = V_{1D}\left(q \sim \frac{1}{d}\right)$, where $d \sim 25$ nm is the distance to the metallic gate. These considerations lead us to estimate $\beta \sim \frac{e}{2\pi\epsilon_0\epsilon\, g\mu_B v_e} \ln\left(\frac{d}{a}\right)$, where $g$ and $v_e$ are the typical values for the $g$-factor and edge speed, at the Fermi wave vector for the edge electrons. For the typical values of $g \sim 5$, $v_F \sim 5 \cdot 10^5$ m/s, and $d/a \sim 5$, we obtain $\beta \sim 0.05$ T/nA. This figure is about an order of magnitude larger than the one needed to fit the data (see Fig. S12). It probably shows that the Coulomb interaction is screened more substantially than implied by the above estimate. Given that the parameters that enter into the estimate for $\beta$ are not well known, one can turn it around to get an estimate for the dimensionless interaction strength on the edge, $V/v_e = \frac{e^2}{2\pi\epsilon_0\epsilon\, \hbar v_e} \ln\left(\frac{d}{a}\right)$ from the measured values of $\beta \sim 3 \cdot 10^{-3}$ T/nA: $V/v_e \sim eg\beta\mu_B/\hbar \sim 0.1$. That is, the edge is not strongly interacting.

SI VIII. Nonlinear measurements

Figure S9 shows two-terminal $I-V$ traces for device MW5 at $T = 5.5$ K. When $\boldsymbol{B}$ is not perpendicular to $\boldsymbol{d}_{so}$ (for example, $B \parallel z$-axis), the $I-V$ trace is different for $+0.3$ T and $-0.3$ T. The current has a component even in $V$ that changes sign when $\boldsymbol{B}$ is reversed, i.e. with the symmetry of $V^2 B$. When $\boldsymbol{B}$ is perpendicular to $\boldsymbol{d}_{so}$, this component is either very small or it vanishes.

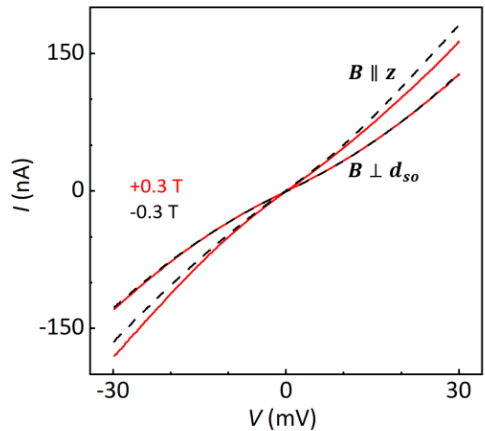

**FIG. S9.** $I - V$ traces at $B = \pm 0.3$ T, $T = 5.5$ K, and $V_g = -2.7$ V for device MW5.

Then an a.c. bias of amplitude $V_f = 15$ mV is applied to measure $G$ and $\gamma$. Figure S10(a) is a MIM image of device MW7. The dependences of $G$ and $\gamma$ on $B$ for contacts 1-2 and 3-4 are plotted in Fig. S10(b) and (c), respectively, when $\boldsymbol{B}$ is applied along the $z$-axis.

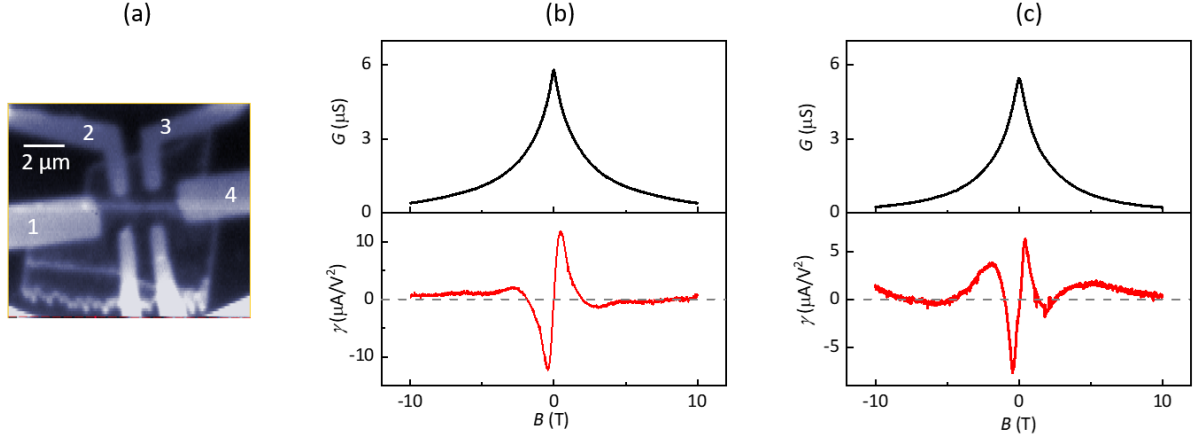

**FIG. S10. Dependence of $G$ and $\gamma$ on $B$ for device MW7.** (a) MIM image of device MW7. Scale bar: 2 μm. (b) and (c) Measurements of the linear conductance, $G = I_{1f}/V_f$ (black), and quadratic coefficient, $\gamma = 2I_{2f}/V_f^2$ (red), vs out-of-plane field, for contacts 1-2 (panel (b)) and 3-4 (panel (c)) in device MW7 ($T = 4$ K $V_g = 0$, $V_f = 15$ mV).

An a.c. bias of amplitude $V_f = 15$ mV is then applied to measure $G = I_f/V_f$ and $\gamma = I_{2f}/2V_f^2$ as a function of $\alpha$ at different $\delta$. Results are shown in Fig. S11. At four different values of $\delta$, $\gamma_a$ always passes through zero when $G$ is minimal.

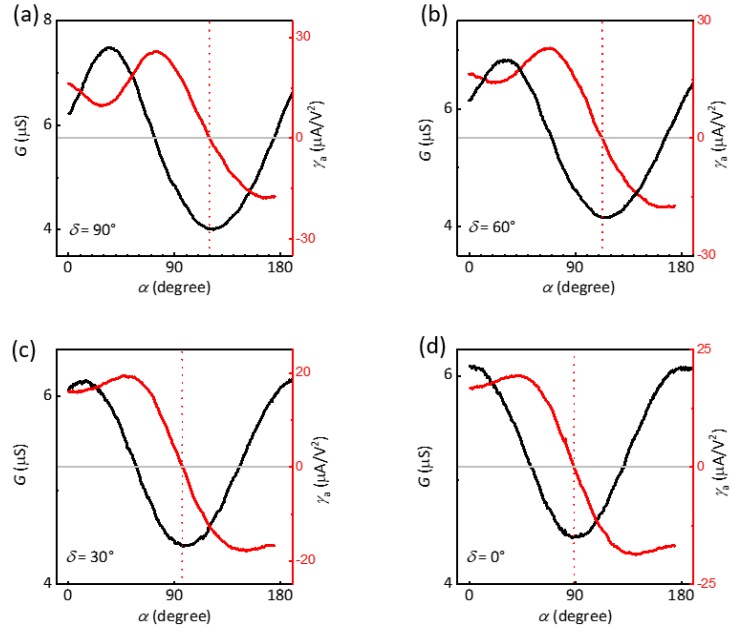

**FIG. S11. Nonlinear measurements at different value of $\delta$ for device MW5.** (a) - (d) First-harmonic response ($G$, black) and second-harmonic response ($\gamma_a$, red) vs. $\alpha$ for $B = 0.3$ T, at $\delta = 90°, 60°, 30°$ and $0°$, respectively. The red dashed lines mark the values of $\alpha$ where $\gamma_a = 0$.

Next, we fit $\gamma$ by measuring the $B$ dependence of $G$ at a series of values of $\varphi$ in the *y-z* plane. Comparison of fitted $\gamma$ (black dashed line) and directly measured $\gamma$ (red dots) is shown in Fig. S12. Equation (4) fits the higher field data in Fig. S12(a) well with $\beta = 3.5 \times 10^{-3}$ T/nA. However, we find that the lower field data can only be fitted well if the second term in equation (4) is neglected, as shown in Fig. S12(b), in which case we get $\beta = 1.0 \times 10^{-3}$ T/nA.

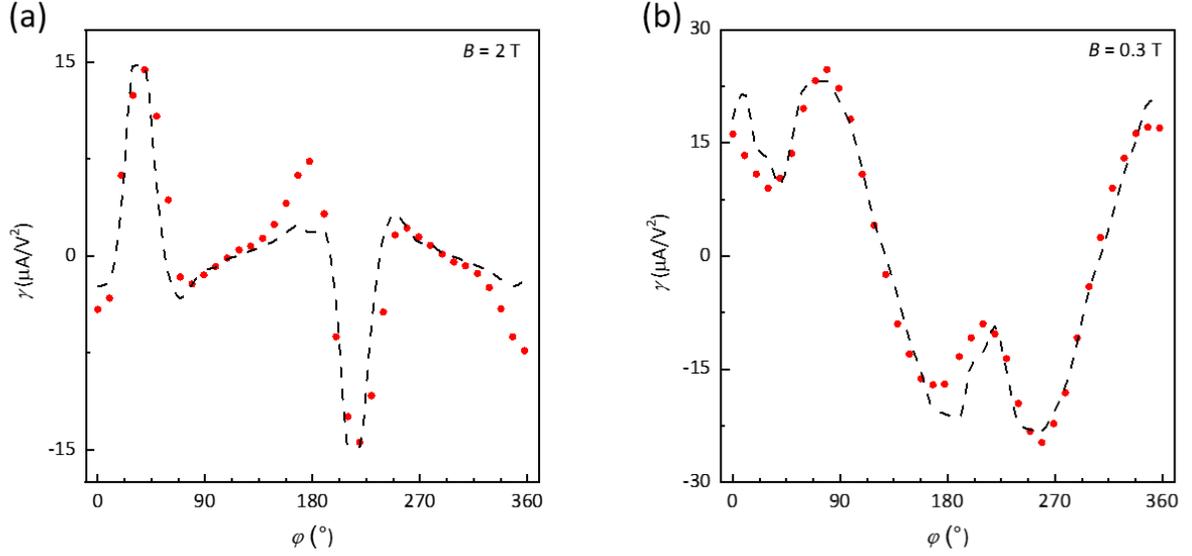

**FIG. S12. Fitting for $\gamma$ in the *y-z* mirror plane for device MW5.** (a) $B = 0.3$ T. (b) $B = 2$ T. Measured and fitted $\gamma$ are shown by red dots and black dashed lines, respectively.

## SI IX. Bilayer WTe$_2$

As additional evidence that the anisotropy of the magnetoconductance and the nonlinear effect comes from the edge of the monolayer WTe$_2$, we present similar measurements made on bilayer WTe$_2$, which lacks edge states (Fig. S13). First, we note that the anisotropy of magnetoconductance shows maxima and minima only at either $\alpha = 0°$ or $90°$ (i.e., perpendicular or parallel to the plane), whatever the temperature and gate voltage. For monolayer WTe$_2$, the conductance shows maxima at $\alpha \approx 40°$ for the value of $\delta$ of Fig. S13(b). Second, the $I - V$ characteristics do not change when the magnetic field is reversed, i.e., $\gamma_a = 0$.

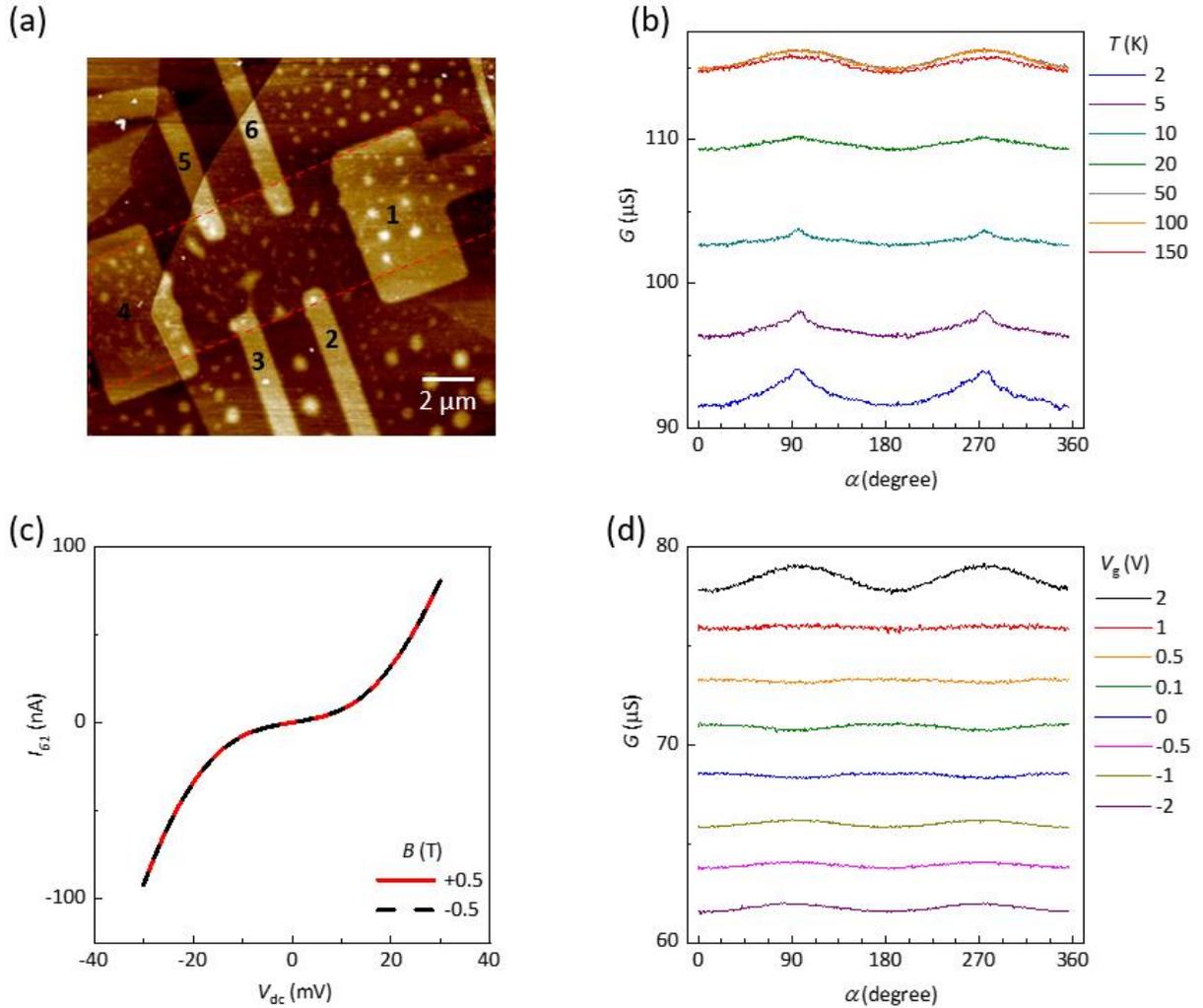

**FIG. S13. Measurement of bilayer WTe$_2$ device BW7.** (a) AFM image of device BW7. The bilayer WTe$_2$ flake is outlined by a red dashed line. Scale bar: 2 µm. (b) Temperature dependence of linear two-terminal conductance on $\alpha$ at $\delta \approx 90°$ and $B = 9$ T and $V_g = 2$ V. The traces are vertically offset for clarity. (c) $I - V$ characteristics at $B = +0.5$ T (red) and $-0.5$ T (black) at $T = 2$ K when magnetic field is out-of-plane. (d) Gate dependence of the linear two-terminal conductance vs $\alpha$ at $\delta \approx 90°$, $B = 9$ T and $T = 50$ K. The traces are vertically offset for clarity.